\documentclass[11pt,twoside]{./atmp}

\usepackage{amsmath,amssymb}
\usepackage{epsfig}

\usepackage[all]{xy}
\usepackage{color}
\newtheorem{Caso}{Case}
\newtheorem{Lema}{Lemma}
\newtheorem{Teorema}{Theorem}

\begin{document}

\title[The equation of eigen-values]
{The equation of eigen-values in a general Euclidean Schwarzschild metric}


\author[Jos\'e L. Mart\'\i nez-Morales]{Jos\'e L. Mart\'\i nez-Morales}

\address{Instituto de Matem\'aticas\\
Universidad Nacional Aut\'onoma de M\'exico\\
A.P. 273, Admon. de correos \#3\\
C.P. 62251 Cuernavaca, Mor.\\
MEXICO}  
\addressemail{martinez@matcuer.unam.mx}

\begin{abstract}
A metric more general than the Euclidean Schwarzschild-Tangherlini metric is considered. The cosmological constant is not necessarily Zero, and the hypersphere is replaced by an Einstein variety.

A differential equation that derives from the equation of eigen-values of the Laplace operator in this metric is studied. Two cases are considered, (i) the cosmological constant is not Zero, and (ii) the cosmological constant is Zero, and the set of eigen-values is discrete.

Infinity is a singular point of this differential equation. In these cases only, it is regular. Series of powers solve the differential equation.

With the eigenvalues from the discrete set, the determinant of the Laplacian is calculated. This is equal to the period of the imaginary time coordinate.

Finally, the stability of the Euclidean Schwarzschild metric is investigated at the classical level.
\end{abstract}

\maketitle
\section{Introduction}
We consider the Klein-Gordon equation in a (1+$n$)-dimensional Euclidean\\Schwarzschild-Tangherlini background. We separate variables and analyze a resulting radial equation and derive recurrence relations for the coefficients of a Taylor expansion.

A more general Schwarzschild-Tangherlini black hole in 1+$n$ space-time dimensions has metric
\begin{center}
-$\left(\frac{{R_h}}{(n-2) (n-1)}-\frac{\rho^{n-2}}{r^{n-2}}-\frac{{r^2} {R_g}}{n (1+n)}\right)dt\sp2$+$\frac{dr\sp2}{\frac{{R_h}}{(n-2) (n-1)}-\frac{\rho^{n-2}}{r^{n-2}}-\frac{{r^2} {R_g}}{n (1+n)}}$+$r\sp2h$,
\end{center}
where $\rho$, $R_h$ and $R_g$ are constants. They are not related to each other. $\rho$ must be a zero of the metric coefficient in front of the $dt^2$ (otherwise it would not make sense to the speak of a ``horizon"). The cosmological constant is not necessarily Zero, and we replace the hypersphere by an Einstein variety ($M$, $h$) of dimension $n$-1. We apply analytic continuation to create a Euclidean (positive definite) signature analogue of the Lorentzian vacuum solution to Einstein's equation \cite{9} \cite{25} \cite{Hu} \cite{Bhawal} \cite{Sorokin} \cite{Galli} \cite{Olesen}\footnote{The reader may be able to track down the reference to the paper by Xing Hu (1999) [in Chinese]. It is fortunate that there seem to be relatively recent references related to the problem under consideration \cite{Hafner} \cite{Hassell} \cite{Corvino}.}:
\[
t\to it.
\]
The metric becomes
\begin{equation}
\label{1}
\hbox{$\left(\frac{{R_h}}{(n-2) (n-1)}-\frac{\rho^{n-2}}{r^{n-2}}-\frac{{r^2} {R_g}}{n (1+n)}\right)dt\sp2+\frac{dr\sp2}{\frac{{R_h}}{(n-2) (n-1)}-\frac{\rho^{n-2}}{r^{n-2}}-\frac{{r^2} {R_g}}{n (1+n)}}+r\sp2h$}.
\end{equation}
The radial variable is now restricted to the range $r\ge \rho$, and regularity at $r$=$\rho$ requires that $t$ be periodic with period 4$\pi \rho$ (this is the general euclidean black hole background). Thus, the space-time asymptotically has one direction compacted on a circle. The obvious symmetry is $U(1)\times SO(n-1)$. The Schwarzschild-Tangherlini solution is invariant under the large symmetry group: $U$(1)$\times SO$($n$). The maximum curvature is of order 1/$\rho\sp2$, which can be made arbitrarily small by taking $\rho$ large. There are no singularities and the space-time is geodesically complete.

This solution describes a contracting and then expanding ``bubble of nothing" in the following sense. Consider the geometry on the $\theta$=$\pi$/2 surface in the Schwarzschild metric, where the parameter $\theta$ is introduced as an angular variable. This resembles the plane {\bf R}$\sp2$ with a circle of radius $\rho$ removed. Over each point is a circle whose radius smoothly goes to zero at $r$=$\rho$ (see figure 1). Thus $r$=$\rho$ is not a boundary of the space, but it is the circle $S\sp1$ of minimal length.
\begin{center}
\includegraphics{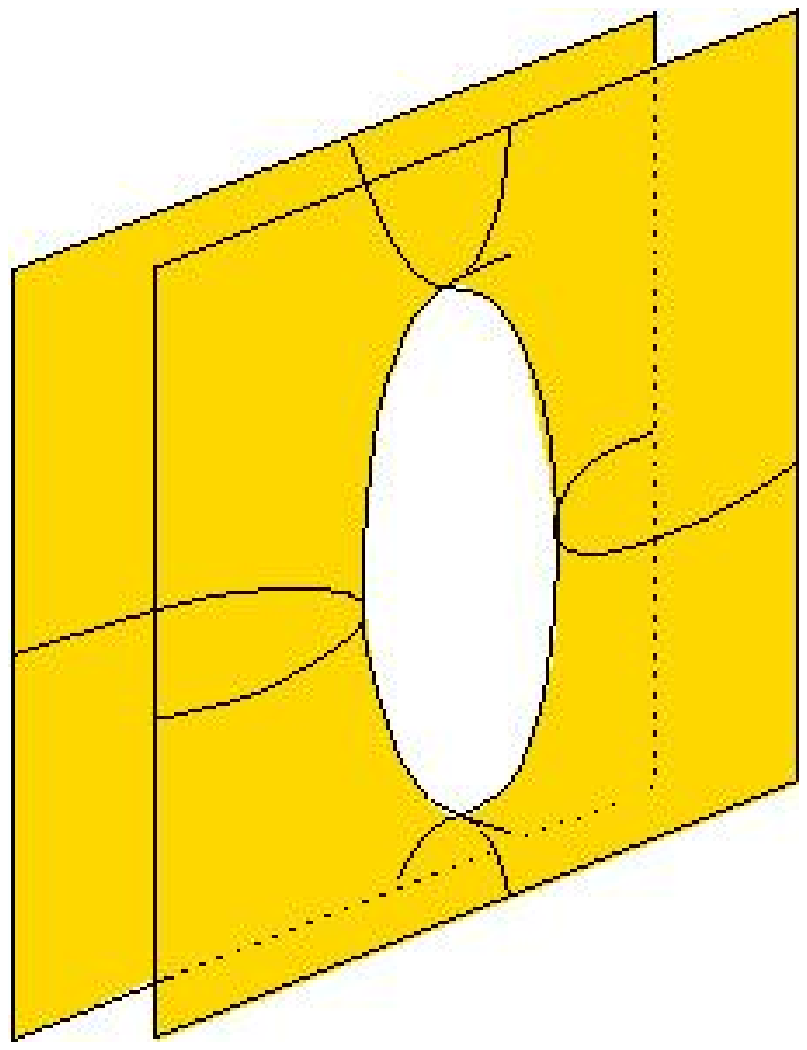}
\end{center}
{\small Figure 1: A schematic depiction of the geometry at a fixed time, with the circle replaced by two points, and the $r$ and $\phi$ directions manifest.}

The reader must be aware that quite a lot work was done in this area in 1980, beyond the references that we
list. In particular, see \cite{Perry} and
other work referred to therein, which covers all the work
of this period on perturbations of the positive-definite Schwarzschild
geometry.\bigskip

We study a differential equation that derives from the equation of eigen-values of the Laplace operator in the metric (\ref{1}). We consider two cases, (i) the cosmological constant is not Zero, and (ii) the cosmological constant is Zero, and the set of eigen-values is discrete.

Infinity is a singular point of this differential equation. In these cases only, it is regular. Series of powers solve the differential equation.\bigskip

The problem studied in this paper is somewhat puzzling. Normally in studying the eigenvalues of the Laplacian on a non-compact manifold, the domain of definition is a crucial point. One can work in the Hilbert space $L^2$; since the manifold is complete, the operator is essentially self-adjoint and the $L^2$ spectrum is well defined. However, in general, the $L^2$ spectrum is continuous and the question of eigenfunctions a difficult one. One can also impose boundary conditions by taking finite limits and then studying the
behavior as the boundaries tend to infinity. 

The current work, however, takes neither of these approaches. We simply postulate a power series expansion, use spherical harmonics, and obtain some results on the power series coefficients. The main theorem is Theorem \ref{Teorema}; one assumes that all the Laurent series coefficients are not necessarily non-zero for index $i\ge$ 0. One then has a recursive relation for the coefficients. The powers of the variable are multiples of a natural number. Therefore, the recursive relation separates into
\begin{itemize}
\item a recursive relation in terms of the eigen-value and the cosmological constant,
\item a recursive relation in terms of the eigen-value on the Einstein variety and its scalar of curvature, and
\item a mixed term that is Zero.
\end{itemize}
Furthermore, it is clear why one would care about having a recursive relation. One must not assume that all the coefficients are non-zero for large $i$. We have a power series in $r^{-n}$, if the cosmological constant is not Zero, and a power series in $r^{2-n}$, if the cosmological constant is Zero.

There is real motivation for the paper. For example, one should only be interested in solving the eigenvalue equation using a separation of variables rather than the more customary methods because if the cosmological constant is Zero, then infinity is a singular point of this differential equation that is regular only on a discrete set of eigen-values. Having results on the eigen-values is important because we can calculate the determinant of the Laplace operator or, in general, we can calculate a partition function in terms of the Laplace operator using these quantized eigen-values.
\section{Classical stability}
In this section we investigate and state our results on stability of the metric (\ref{1}) under perturbations of the metric at the classical level. These satisfy the scalar wave equation. We are required to check whether there are
normalizable effectively tachyon modes (namely, modes which grow in time relative to
the background metric) among the solutions to the linearized field equations. Such modes
would be localized near the bubble since the asymptotic region has no such excitations.

We begin by considering a scalar field $TRF$. In this background, it is in the kernel of the following operator:
\begin{center}
-$\frac1{\frac{{R_h}}{(n-2) (n-1)}-\frac{\rho^{n-2}}{r^{n-2}}-\frac{{r^2} {R_g}}{n (1+n)}}\frac{\partial\sp2}{\partial t\sp2}$-${r^{1-n}}\frac\partial{\partial r} \bigg(\frac{{r^{n-1}} {R_h}}{(n-2) (n-1)}$-$r {{\rho}^{n-2}}$-$\frac{{r^{1+n}} {R_g}}{n (1+n)}\bigg)\frac\partial{\partial r}$+$\frac{{{\Delta}_h}}{{r^2}}$-$\lambda$,
\end{center}
where $\lambda$ is the bare mass.

We denote by $\Delta_h$ the Laplace operator in the metric $h$. Let us separate variables and consider modes with definite angular momentum $\mu$, $\Delta_hF$=$\mu F$, and definite momentum $\sqrt{\nu}$ around the $t$ direction, $\partial_t\sp2T$=-$\nu T$.
\begin{Lema}
\begin{equation}
\label{3}
\frac{\frac{\mu}{{r^2}}+\frac\nu{\frac{{R_h}}{(n-2) (n-1)}-\frac{\rho^{n-2}}{r^{n-2}}-\frac{{r^2} {R_g}}{n (1+n)}}-\lambda}{\frac{\rho^{n-2}}{r^{n-2}}+\frac{{r^2} {R_g}}{n (1+n)}-\frac{{R_h}}{(n-2) (n-1)}}R+\frac{\frac{\rho^{n-2}}{r^{n-1}}+\frac{r {R_g}}{n}-\frac{{R_h}}{(n-2) r}}{\frac{\rho^{n-2}}{r^{n-2}}+\frac{{r^2} {R_g}}{n (1+n)}-\frac{{R_h}}{(n-2) (n-1)}}{R^{\prime}}+{R''}=0.
\end{equation}
\end{Lema}

For now let us just consider modes with $\nu$=0 and particles with non zero definite angular momentum, $\mu\neq$ 0.

Consider a natural number $l$. The solutions for the Schwarzschild metric are as follows. Write
\[
\mu=l(1+l).
\]
We have polynomial solutions
\[
R\sim\sum _{i=0}^{l}\bigg(\prod _{j=1}^{i}\frac{{{(1-j+l)}^2}}{-1+j-2 l}\bigg)\frac{{(\frac r\rho)^{l-i}}}{i!}.
\]
This generalizes the metric perturbations of the Schwarzschild black hole studied explicitly in \cite{27} \cite{28} \cite{WaldErratum} \cite{Wald}, at least for $\nu$=0 (higher $\nu$ modes might be expected to be less tachyon in any case). One can translate the modes studied in \cite{27} \cite{28} \cite{WaldErratum} \cite{Wald}, expressed in terms of tensor spherical harmonics in the black hole background, to modes in the Euclidean Schwarzschild geometry. In the former case, the spherical harmonics are eigenfunctions of the Laplacian on the spherical directions of the black hole with eigenvalue $l$($l$ + 1); in our analytic continuation these become tensor spherical harmonics on slices of (\ref{1}) with $\mu$=$l$($l$ + 1). The time direction in the black hole becomes our $t$ direction, and imaginary frequency for the black hole modes (corresponding to tachyons in that geometry) would correspond to real momentum $\sqrt{\nu}$ along our $t$ circle, which for us is quantized. The analysis of \cite{27} \cite{28} \cite{WaldErratum} \cite{Wald} rules out tachyons in the black hole background by showing that there are no normalizable and nonsingular solutions to the radial part of the equations of motion, and this directly rules out tachyon modes in the bubble of $\mu$=$l$($l$+1). Furthermore, we find that by replacing $l$($l$+1) by $\mu$ in their analysis, and rescaling fields appropriately, one can extend their arguments to arbitrary $\mu$, at least for $\nu$=0. Thus, at least in four dimensions, (\ref{1}) appears to be a classically stable solution to the equations of motion.
\section{The differential equation in the radial variable}
We study the differential equation (\ref{3}) in the radial variable that derives from the equation of eigen-values of the Laplace operator. We consider two cases, (i) the cosmological constant is not Zero, and (ii) the cosmological constant is Zero, the dimensions are high, and the set of eigen-values is discrete.

Infinity is a singular point of the differential equation. In these cases only, it is regular.

Series of powers solve the differential equation. The coefficients of the series fulfill a recursive relation in terms of the space dimension,
\begin{itemize}
\item the cosmological constant, the eigen-value and the square of the radius of the horizon, in the first case,
\item the scalar of curvature of the variety, in the second case.
\end{itemize}
We then study the static basis state on the Euclidean Schwarzschild-\\Tangherlini metric. In the space-time of four dimensions, logarithms and polynomials solve the differential equation.\bigskip

\begin{Lema}
Consider the function $f$ on (0, 1) defined as the function $R$ evaluated in the product of the radius $\rho$ and the inverse $x\sp{-1}$ of the variable,
\[
f(x)=R(\rho x\sp{-1}).
\]
Then, it solves a differential equation,
\begin{equation}
\label{4}
\hbox{$f''=\rho^2\frac{\lambda-\frac{{x^2} \mu}{{{\rho}^2}}-\frac{\nu}{\frac{{R_h}}{(n-2) (n-1)}-{x^{n-2}}-\frac{{{\rho}^2} {R_g}}{n (1+n) {x^2}}}}{{x^{2+n}}+\frac{{x^2} {{\rho}^2} {R_g}}{n (1+n)}-\frac{{x^4} {R_h}}{(n-2) (n-1)}}f-\frac{{x^n}+\frac{(1-n){{\rho}^2} {R_g}}{n (1+n)}+\frac{(n-3) {x^2} {R_h}}{(n-2) (n-1)}}{{x^{1+n}}+\frac{{x} {{\rho}^2} {R_g}}{n (1+n)}-\frac{{x^3} {R_h}}{(n-2) (n-1)}}{f^{\prime}}$}.
\end{equation}
\end{Lema}
Consider two cases.
\begin{Caso}
\label{Caso 1}
The scalar of curvature $R_g$ of the Cartesian product is not Zero.
\end{Caso}
\begin{Caso}
\label{Caso 2}
$\hbox{}$
\begin{itemize}
\item The space dimension $n$ is greater than Three,
\item the scalar of curvature $R_g$ of the Cartesian product is Zero, and
\item the eigen-value $\lambda$ of the Laplace operator on the Cartesian product is the quotient of (i) the eigen-value $\nu$ on the circle, and (ii) the sectional curvature $\frac{{R_h}}{(n-2) (n-1)}$ of the variety,
\begin{equation}
\label{}
\lambda=\frac{(n-2) (n-1) \nu}{{R_h}}.
\end{equation}
\end{itemize}
\end{Caso}
\subsection{The determinant of the Laplacian}
In this subsection the determinant of $\Delta$ is computed. The result looks interesting at first sight.

In functional integrals, the regularized determinant of the Laplacian $\Delta$ can be defined in terms of the $\zeta$-function of $\Delta$ by the formula ln det $\Delta$=-$\zeta$'(0). This definition can be used whenever the $\zeta$-function $\zeta$($s$)=$\sum\lambda_i\sp{-s}$ can be analytically continued to the point $s$=0. We assume specific values for the eigenvalues $\lambda_i$.

The correct values for the $\lambda_i$'s are given by
\[
\lambda_i=\frac{(n-2) (n-1)}{{R_h}}\left(\frac{i}{2\rho}\right)\sp2.
\]
Therefore,
\begin{equation}
\label{2}
\zeta(s)=\sum_i\left(\frac{(n-2) (n-1)(\frac{i}{2\rho})\sp2}{{R_h}}\right)\sp{-s}.
\end{equation}
Differentiating (\ref{2}) with respect to $s$ and setting $s$=0, we get
\[
\zeta'(0)=\ln\left(\frac{1}{4 \pi {\rho\sqrt{\frac{R_h}{(n-2)(n-1)}}}}\right).
\]
Therefore,
\[
\det\Delta=4 \pi {\rho\sqrt{\frac{R_h}{(n-2)(n-1)}}}.
\]
If the sectional curvature of the variety is One, then,
\[
\det\Delta=4 \pi {\rho}.
\]
\subsection{The characteristic equation of the differential equation} 
Infinity is a singular point of the differential equation in the radial variable, it is regular only in the cases \ref{Caso 1} or \ref{Caso 2}. We solve the characteristic equation of the differential equation. \bigskip

Set
\begin{eqnarray*}
l_1&=&\lim_{x\to0}\frac{x\bigg({x^{1+n}}+\frac{(1-n) x {{\rho}^2} {R_g}}{n (1+n)}+\frac{(n-3) {x^3} {R_h}}{(n-2) (n-1)}\bigg)}{\bigg({x^{2+n}}+\frac{{x^2} {{\rho}^2} {R_g}}{n (1+n)}-\frac{{x^4} {R_h}}{(n-2) (n-1)}\bigg)},\\
l_2&=&\lim_{x\to0}\frac{x\sp2\Bigg(\frac{{x^2} \mu}{{{\rho}^2}}+\frac{\nu}{\frac{{R_h}}{(n-2) (n-1)}-{x^{n-2}}-\frac{{{\rho}^2} {R_g}}{n (1+n) {x^2}}}-\lambda\Bigg)}{\frac1{\rho^2}\bigg({x^{2+n}}+\frac{{x^2} {{\rho}^2} {R_g}}{n (1+n)}-\frac{{x^4} {R_h}}{(n-2) (n-1)}\bigg)}.
\end{eqnarray*}
\begin{Lema}
The first limit depends on the space dimension,
\[
l_1=1-n+\left\{\begin{array}{ll}
0,&\hbox{Case }\ref{Caso 1}\\
2,&\hbox{Case }\ref{Caso 2}
\end{array}\right..
\]
The second limit is an integer multiple of quotients of eigen-values and scalars of curvature,
\[
l_2=\left\{\begin{array}{ll}
-\frac{n(1+n) \lambda}{{R_g}},&\hbox{Case }\ref{Caso 1}\\
-\frac{216 {\nu} {{\rho}^2}}{R_{h}^{3}}\delta_4\sp n-\frac{(n-2)(n-1) \mu}{{R_h}},&\hbox{Case }\ref{Caso 2}
\end{array}\right..
\]
\end{Lema}
Set
\[
e=\left\{\begin{array}{ll}
\frac{n}{2}\pm\frac{{\sqrt{n}} {\sqrt{4 (1+n) \lambda +n {R_g}}}}{2 {\sqrt{{R_g}}}},&\hbox{Case }\ref{Caso 1}\\
\frac{1}{2} \Bigg(n-2\pm\frac{{\sqrt{n-2}} {\sqrt{4 (n-1) \mu +\frac{432 {\nu} {{\rho}^2}}{R_{h}^{2}}\delta_4\sp n+(n-2) {R_h}}}}{{\sqrt{{R_h}}}}\Bigg),&\hbox{Case }\ref{Caso 2}
\end{array}\right..
\]
\begin{Lema}
$e$ solves the characteristic equation of the differential equation,
\[
{e^2}+({l_1}-1)e+{l_2}=0.
\]
\end{Lema}
\subsection{Solution in series of powers}
If a series of powers $x\sp e\sum_{i=0}\sp\infty a_ix\sp i$ solves the differential equation, then the coefficients of the Taylor expansion fulfill a recurrence relation in terms of
\begin{itemize}
\item the characteristic exponent,
\item the eigen-values,
\item the scalars of curvature,
\item the space dimension, and
\item the square of the radius of the horizon,
\end{itemize}
\[
{(e+i)}^2a_i=
\]
\begin{eqnarray}
&&{{\rho}^2}\Big(\lambda -\frac{(e+i+n) (2(e+i)+n) {R_g}}{ n (1+n)}\Big){a_{i+n}}\nonumber\\
&&+\frac{ {{\rho}^4} {R_g}}{n (1+n)}\Big(\lambda -\frac{(e+i+n) (e+i+2 n) {R_g}}{ n (1+n)}\Big){a_{i+2n}}\nonumber\\
&&-\Big(\mu-\frac{(e+i+(n-2)) (2(e+i)+(n-2)) {R_h}}{ (n-2) (n-1)}\Big){a_{i+n-2}}\nonumber\\
&&+\frac{ {R_h}}{(n-2) (n-1)}\Big(\mu -\frac{(e+i+(n-2)) (e+i+2(n-2)) {R_h}}{ (n-2) (n-1)}\Big){a_{i+2(n-2)}}\nonumber\\
&&+{{\rho}^2} \Big(\nu -\frac{\mu {R_g}}{n (1+n)}-\frac{\lambda {R_h}}{(n-2) (n-1)}\nonumber\\
\label{5}
&&+\frac{2 (e+i+(n-1)) (e+i+2 (n-1)) {R_g}{R_h}}{(n-2) (n-1) n (1+n)}\Big){a_{i+2 (n-1)}}.
\end{eqnarray}
This gives a formal power series for the solutions.

The powers of the variable $x$ are multiples of a natural number. Therefore, the recursive relation separates into
\begin{itemize}
\item a recursive relation in terms of the eigen-value $\lambda$ and the scalar of curvature $R_g$,
\item a recursive relation in terms of the eigen-value $\mu$ and the scalar of curvature $R_h$, and
\item a mixed term (\ref{5}) that is Zero.
\end{itemize}
\begin{Teorema}
\label{Teorema}
Consider a sequence of numbers $a_0$, $a_1$..., that fulfill a recursive relation in terms of the characteristic exponent, the eigen-values, the scalars of curvature and the space dimension,
\[
{(e+i)}^2a_i=\left\{\right.
\]
${{\rho}^2}\Big(\lambda$-$\frac{(e+i+n) (2(e+i)+n) {R_g}}{ n (1+n)}\Big){a_{1+i}}$+$\frac{ {{\rho}^4} {R_g}}{n (1+n)}\Big(\lambda$-$\frac{(e+i+n) (e+i+2 n) {R_g}}{ n (1+n)}\Big){a_{2+i}}$,

\noindent{\footnotesize$\Big(\frac{(e+i+(n-2)) (2(e+i)+(n-2)) {R_h}}{ (n-2) (n-1)}$-$\mu\Big){a_{1+i}}$+$\frac{ {R_h}}{(n-2) (n-1)}\Big(\mu$-$\frac{(e+i+(n-2)) (e+i+2(n-2)) {R_h}}{ (n-2) (n-1)}\Big){a_{2+i}}$},\bigskip

\noindent according to the case. Then, the product of (i) the variable raised to the characteristic exponent $x\sp e$, and (ii) a series of powers multiple of a number that depends on the space dimension,
\begin{equation}
\label{7}
x\sp e\sum_{i=0}\sp\infty a_ix\sp{im},
\end{equation}
solves the differential equation (\ref{4}), where
\[
m=\left\{\begin{array}{l}
n\\
n-2
\end{array}\right.,
\]
according to the case.
\end{Teorema}
Using the formula of iteration satisfied by the sequence $\{a_i\}$, we calculate an equivalent expression of (\ref {7}) in Appendix \ref{Seccion}.
\subsection{Basic state of the Euclidean Schwarzschild Tangherlini metric}
As one result of this paper, we manage to solve the recurrence relations in special cases to obtain solutions as explicit power series or even polynomials. This is interesting.

We denote by $\Gamma$ the Gamma function.
\begin{Lema}
If it is assumed that (i) the eigen-values $\lambda$, $\nu$ and the scalar of curvature $R_g$ vanish, and (ii) the variety is the hypersphere,
\begin{eqnarray*}
{R_h}&=&(n-2) (n-1),\\
\mu &=&l(l+n-2),
\end{eqnarray*}
then two series of powers
\[
{x^{-l}}\sum _{i=0}^{\infty}\frac{\Gamma(-\frac{2 l}{n-2}){{\Gamma(i+\frac{l}{2-n})}^2}}{\Gamma(i-\frac{2 l}{n-2}) {{\Gamma(\frac{l}{2-n})}^2}}\frac{{x^{i(n-2)}}}{i!}
\]
and
\begin{equation}
\label{8}
{x^{n-2+l}}\sum _{i=0}^{\infty}\frac{\Gamma(2+\frac{2 l}{n-2}){{\Gamma(i+1+\frac{l}{n-2})}^2}}{\Gamma(i+2+\frac{2 l}{n-2}) {{\Gamma(1+\frac{l}{n-2})}^2}}\frac{{x^{i(n-2)}}}{i!},
\end{equation}
solve the differential equation (\ref{4}). 
Moreover, in the space-time of four dimensions,
\[
n=3,
\]
\begin{itemize}
\item a rational function
\begin{equation}
\label{9}
{x^{-l}}\sum _{i=0}^{l}\bigg(\prod _{j=1}^{i}\frac{{{(1-j+l)}^2}}{-1+j-2 l}\bigg)\frac{{x^i}}{i!},
\end{equation}
and
\item the product of a rational function and a logarithmic function
\begin{equation}
\label{10}
-(-1)^l\frac{(1+2l)!}{{{l!}^2}}\sum _{i=0}^{l}(-1)^i\frac{(l+i)!}{{{i!}^2}(l-i)!}{x^{-i}}\left(\log(1-x)+\sum _{j=1}^{l+i}\frac{{x^j}}{j}\right)
\end{equation}
solve the differential equation (\ref{4}).
\end{itemize}
\end{Lema}
If
\[
l=1\quad\hbox{and}\quad n=3,
\]
then, the power series in (\ref {8}) becomes
\begin{eqnarray*}
&&6\sum _{i=0}^{\infty }\frac{ (1+i) {x^{2+i}}}{(2+i) (3+i)}\\
&=&6\sum _{i=0}^{\infty }\frac{ (1+i) }{ (3+i)}\int _{0}^{x}{y^{1+i}}d y\\
&=&6\sum _{i=0}^{\infty }\frac{ 1 }{ (3+i)}\int _{0}^{x}y{\frac\partial{\partial y}} {y^{1+i}}d y\\
&=&6\sum _{i=0}^{\infty }\frac{ 1 }{ (3+i)}\Big({x^{2+i}}-\int _{0}^{x} {y^{1+i}}d y\Big)\\
&=&6\bigg(\frac{1}{x}\sum _{i=0}^{\infty }\frac{x^{3+i}}{ (3+i)}-\int _{0}^{x}\frac{1}{{y^2}}\sum _{i=0}^{\infty }\frac{y^{3+i}}{ (3+i)} d y\bigg)\\
&=&6\bigg(\frac{1}{x}\bigg(-x-\frac{{x^2}}{2}-\log(1-x)\bigg)-\int _{0}^{x}\frac{1}{{y^2}}\bigg(-y-\frac{{y^2}}{2}-\log(1-y)\bigg)d y\bigg)\\
&=&6\Big(-2+\Big(1-\frac{2}{x}\Big) \log(1-x)\Big).
\end{eqnarray*}
With this argument we obtain (\ref {10}).

Our solutions which describe the ground state of the system are shown.
\begin{center}
\includegraphics{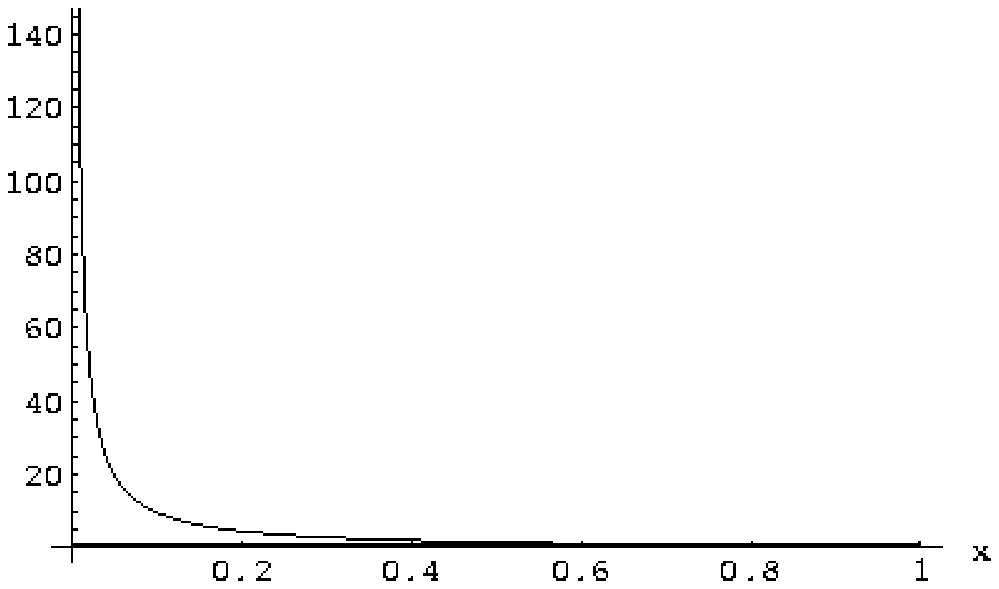}

Plot of (\ref{9}) for $l$=1.\bigskip

\includegraphics{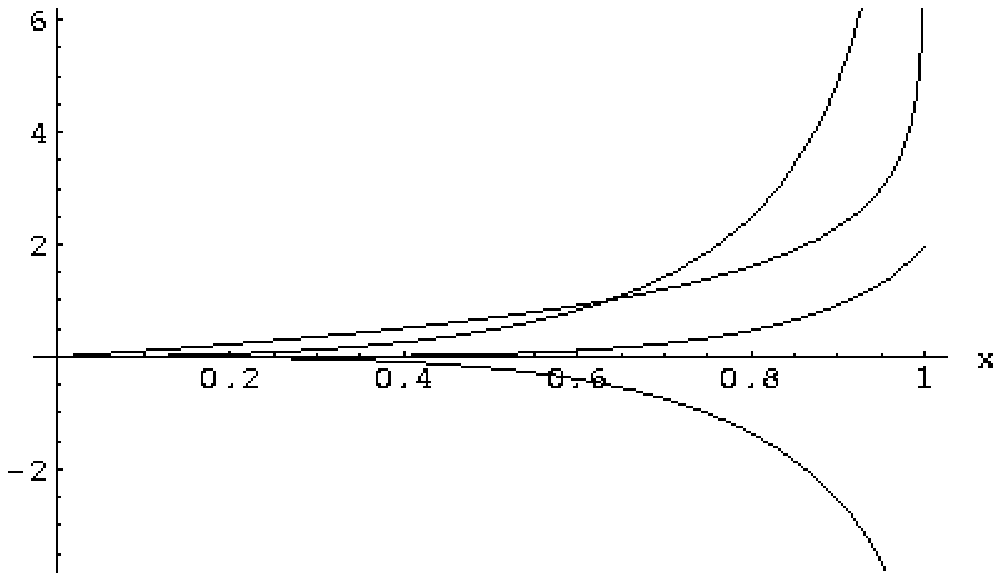}

Plot of (\ref{10}) for $l$=0, 1, 2, 3.
\end{center}
\begin{appendix}
\section{}
\label{Seccion}
Using the formula of iteration satisfied by the sequence $\{a_i\}$, we calculate an equivalent expression of (\ref {7}).

Set
\begin{eqnarray*}
q&=&\left\{\begin{array}{l}
\frac{n(1+n)}{\rho\sp2R_g}\\
\frac{(n-2)(n-1)}{R_h}
\end{array}\right.,\\
b_i&=&\left\{\begin{array}{l}
\frac{n(1+n)}{\rho\sp4R_g}\left[\frac\lambda{(e+i)\sp2}-\frac{R_g}{n(1+n)}\left(1+\frac n{e+i}\right)\left(1+2\frac n{e+i}\right)\right]\sp{-1}\\
\frac{(n-2)(n-1)}{R_h}\left[\frac\mu{(e+i)\sp2}-\frac{R_h}{(n-2)(n-1)}\left(1+\frac{n-2}{e+i}\right)\left(1+2\frac{n-2}{e+i}\right)\right]\sp{-1}
\end{array}\right.,
\end{eqnarray*}
according to the case. Then,
\[
a_{2+i}=qa_{1+i}+a_ib_i.
\]
For a pair of natural numbers $i$ and $j$, we define the set $S_ {i, j} $ of functions $f$: $ \{$1, ..., $i \} \to \{$1, ..., $j \} $ such that
\[
f(1)=1,\quad f(i)=f(j)\quad\hbox{and}\quad1+f(k)<f(1+k).
\]
Consider the numbers
\begin{eqnarray*}
a_{i,j}=\sum_{f\in S_{i,2i+j-1}}\prod_{k=1}\sp i b_{f(k)},\\
b_{i,j}=\sum_{f\in S_{i,2i+j-1}}\prod_{k=1}\sp i b_{1+f(k)}.
\end{eqnarray*}
If
\[
a_0=0\quad\hbox{and}\quad a_1=1,
\]
then,
\[
\sum_{i=0}\sp\infty a_ix\sp i=
\]
\[
x \big(1+q x+{q^2} {x^2}+ ...\big)+{x^3} \big(1+q x+{q^2} {x^2}+ ...\big) {b_1}+q {x^4} \big(1+q x+{q^2} {x^2}+ ...\big) {b_2}+...
\]
\[
{x^5} \big(1+q x+{q^2} {x^2}+ ...\big) {b_1} {b_3}+q {x^6} \big(1+q x+{q^2} {x^2}+ ...\big) {b_1} {b_4}+q {x^6} \big(1+q x+ ...\big) {b_2} {b_4}+...
\]
\[
{x^7} \big(1+q x+{q^2} {x^2}+ ...\big) {b_1} {b_3} {b_5}+q {x^8}\big(1+q x+ ...\big) {b_1} {b_3} {b_6}+q {x^8}\big(1+q x+ ...\big) {b_1} {b_4} {b_6}+...
\]
\[
=\frac x{1-qx}\left(1+\sum_{i=1}\sp\infty\sum_{j=0}\sp\infty a_{i,j}q\sp jx\sp{2i+j}\right).
\]
Similarly, if
\[
a_0=1\quad\hbox{and}\quad a_1=0,
\]
then,
\[
\sum_{i=0}\sp\infty a_ix\sp i=1+\frac{b_0x\sp2}{1-qx}\left(1+\sum_{i=1}\sp\infty\sum_{j=0}\sp\infty b_{i,j}q\sp jx\sp{2i+j}\right).
\]
\end{appendix}

\bibliographystyle{my-h-elsevier}

\end{document}